\documentclass[journal]{IEEEtran}
\usepackage{amsmath,amsfonts}
\usepackage[ruled,linesnumbered]{algorithm2e}
\usepackage{array}
\usepackage[caption=false,font=normalsize,labelfont=sf,textfont=sf]{subfig}
\usepackage{textcomp}
\usepackage{stfloats}
\usepackage{url}
\usepackage{verbatim}
\usepackage{graphicx}
\usepackage{cite}
\usepackage{tabularx}
\usepackage{hyperref} 
\usepackage{threeparttable}
\usepackage{multirow}
\usepackage{colortbl}
\usepackage{orcidlink}
\newcolumntype{P}[1]{>{\centering\arraybackslash}p{#1}}

\begin{document}

\title{FireFly-S: Exploiting Dual-Side Sparsity for Spiking Neural Networks Acceleration with Reconfigurable Spatial Architecture}

\author{
  Tenglong Li \orcidlink{0009-0007-3266-2075},
  Jindong Li \orcidlink{0000-0002-4009-916X},
  Guobin Shen \orcidlink{0000-0002-4069-2107},
  Dongcheng Zhao \orcidlink{0000-0002-0593-8650}, 
  Qian Zhang \orcidlink{0000-0001-5314-4233},
  Yi Zeng \orcidlink{0000-0002-9595-9091}

  \thanks{Manuscript created 29 August 2024. This work was supported by the Chinese Academy of Sciences Foundation Frontier Scientific Research Program (ZDBS-LY-JSC013). \textit{(Corresponding authors: Qian Zhang; Yi Zeng.)}}

  \thanks{Tenglong Li, Jindong Li, and Qian Zhang are with the School of Artificial Intelligence, University of Chinese Academy of Sciences, Beijing 100049, China, and also with the Brain-inspired Cognitive Intelligence Lab, Institute of Automation, Chinese Academy of Sciences, Beijing 100190, China (e-mail: litenglong2023@ia.ac.cn, lijindong2022@ia.ac.cn, q.zhang@ia.ac.cn).}
  
  \thanks{Guobin Shen is with the School of Future Technology, University of Chinese Academy of Sciences, Beijing 100049, China, and also with the Brain-inspired Cognitive Intelligence Lab, Institute of Automation, Chinese Academy of Sciences, Beijing 100190, China (e-mail: shenguobin2021@ia.ac.cn).}
  
  \thanks{Dongcheng Zhao is with the Brain-inspired Cognitive Intelligence Lab, Institute of Automation, Chinese Academy of Sciences, Beijing 100190, China (e-mail: zhaodongcheng2016@ia.ac.cn).}  
  
  \thanks{Yi Zeng is with the Brain-inspired Cognitive Intelligence Lab, Institute of Automation, Chinese Academy of Sciences, Beijing 100190, China, and University of Chinese Academy of Sciences, Beijing 100049, China, and also with the Key Laboratory of Brain Cognition and Brain-inspired Intelligence Technology, Chinese Academy of Sciences, Shanghai 200031, China (e-mail: yi.zeng@ia.ac.cn).}
}

\markboth{IEEE Transactions on Circuits and Systems---I: Regular Papers,~Vol.~72, No.~8, August~2025}%
{Li \MakeLowercase{\textit{et al.}}: FireFly-S: Exploiting Dual-Side Sparsity for Spiking Neural Networks Acceleration}


\maketitle

\begin{abstract}
Spiking Neural Networks (SNNs), with their brain-inspired structure using discrete spikes instead of continuous activations, are gaining attention for their potential of efficient processing on neuromorphic chips.
While current SNN hardware accelerators often prioritize temporal spike sparsity, exploiting sparse synaptic weights offers significant untapped potential for even greater efficiency.
To address this, we propose FireFly-S, a Sparse extension of the FireFly series. This co-optimized software-hardware design focuses on leveraging dual-side sparsity for acceleration.
On the software side, we propose a novel algorithmic optimization framework that combines gradient rewiring for pruning and modified Learned Step Size Quantization (LSQ) tailored for SNNs, achieving a remarkable weight sparsity exceeding 85\% and enabling efficient 4-bit quantization with negligible accuracy loss.
On the hardware side, we present an efficient dual-side sparsity detector employing a Bitmap-based sparse decoding logic to pinpoint the positions of non-zero weights and input spikes. The logic allows for direct bypassing of redundant computations, thereby enhancing computational efficiency.
Different from the overlay architecture adopted by previous FireFly series, we adopt a parametric spatial architecture with inter-layer pipelining that can fully exploit the fine-grained programmability and reconfigurability of Field-Programmable Gate Arrays (FPGAs), enabling fast deployment for various models. A spatial-temporal dataflow is also proposed to support such inter-layer pipelining and avoid long-term temporal dependencies.
In experiments conducted on the MNIST, DVS-Gesture and CIFAR-10 datasets, the FireFly-S model achieves 85--95\% sparsity with 4-bit quantization and the hardware accelerator effectively leverages the dual-side sparsity, delivering outstanding performance metrics of 10,047~FPS on MNIST, 3,683~FPS on DVS-Gesture, and 2,327~FPS on CIFAR-10.

\end{abstract}


\section{Introduction}
\IEEEPARstart{A}{s} a promising alternative to Artificial Neural Networks (ANNs), Spiking Neural Networks (SNNs) offer compelling advantages due to their high biological plausibility, effective event-driven processing, and inherent energy efficiency \cite{maass1997networks}. Moreover, spurred by both biological discoveries and advancements in deep learning, recent SNN algorithms have significantly closed the performance gap with ANNs \cite{shen2022backpropagation}. However, real-world deployment of SNNs on hardware poses unique challenges stemming from their inherent complexity. In contrast to the dense, frame-based dataflow of ANNs, SNNs, characterized by spikes and time steps, exhibit a sparse and temporally correlated dataflow.

Despite the parallel processing capabilities of graphics processing units (GPUs), which are frequently used to deploy SNNs, their architecture is not tailored to the unique characteristics of SNNs. Specifically, GPUs are not designed for event-driven processing and are not optimized for the 1-bit arithmetic operations commonly found in SNNs, highlighting the need for dedicated hardware solutions. In response to these shortcomings, researchers have explored the use of application-specific integrated circuits (ASICs) and field-programmable gate arrays (FPGAs). Notable developments include IBM's TrueNorth chip \cite{akopyan2015truenorth} and Intel's Loihi \cite{davies2018loihi} and its successor Loihi2 \cite{orchard2021efficient}, which are ASICs designed explicitly for SNNs and demonstrate marked improvements in energy efficiency and event-based processing. FPGAs also play a crucial role in SNNs acceleration, offering a flexible platform for exploring various optimization strategies. Some designs leverage this flexibility to enhance both arithmetic and memory efficiency, as in our previous work, FireFly \cite{li2023firefly} and FireFly v2 \cite{li2024firefly}, while others focus on exploiting the sparse nature of spikes through event-driven processing \cite{panchapakesan2022syncnn} \cite{liu2022fpga}.

However, although these designs have made significant strides in improving the performance and energy efficiency of SNN hardware accelerators, they often overlook the potential for further optimization in areas such as network topology and synaptic connectivity. Exploiting the dual-side sparsity, specifically synaptic weight sparsity and activation spike sparsity, can substantially reduce both computational load and network parameters. Moreover, their reliance on high-precision data types like FP32, FP16 for weights and membrane potentials \cite{liu2022fpga} \cite{chen2024sibrain}, also increases the demand for DSP resources and large memory footprints, hindering deployment and limiting efficiency on resource-constrained platforms.

Based on these observations, we introduce FireFly-S, a co-optimized software-hardware design that capitalizes on dual-side sparsity for enhanced acceleration, as illustrated in Fig. \ref{fig:system}. Our software stack initiates by pruning and quantizing the network weights during training, thereby inducing weight sparsity and low-bit representation to diminish both the model size and computational requirements. Subsequently, our hardware implementation maps the optimized SNN model onto FPGA fabric in a fully pipelined fashion. Each layer is equipped with a dedicated dataflow orchestrator that manages inter-layer data transfer, while parallel sparse detectors effectively leverage the dual-side sparsity of both sparse weights and spikes to optimize neuronal dynamics within each layer. Our contributions can be summarized as follows:

\begin{figure*}
  \centering
  \includegraphics[width=1.0\linewidth]{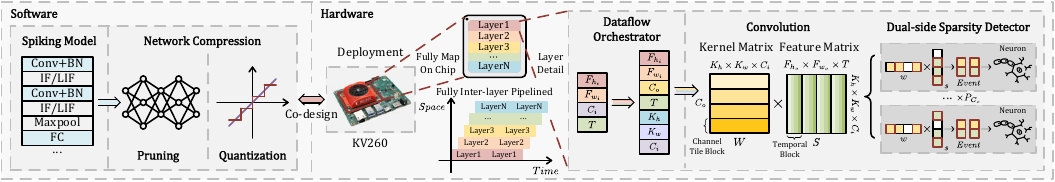}
  \caption{
    Overview of the FireFly-S system architecture, combining the software optimization framework and hardware accelerator.
  }
  \label{fig:system}
\end{figure*}


\begin{enumerate}
  \item {
    FireFly-S proposes an algorithmic optimization framework that combines gradient rewiring-based pruning and a Learned Step Size Quantization (LSQ) variant specifically tailored for SNNs, achieving over 85\% weight sparsity with 4-bit quantization during training for various SNN models.
  }
  \item {
    FireFly-S employs a Bitmap-based sparse decoding logic in the dual-side sparsity detector, enabling efficient exploitation of sparsity found in both activation spikes and synaptic weights. This approach directly skips all computations involving zero values, eliminating the inefficiency of performing unnecessary and idle calculations.
  }
  \item {
    FireFly-S leverages a fully unfolded spatial architecture with inter-layer pipelining, eliminating off-chip memory access and concurrently processing multiple layers. We also present an inter-layer dataflow tailored for SNNs, along with a dedicated logic to bridge the dataflow between layers, minimizing the need for extensive on-chip buffering.
  }
\end{enumerate}

The remaining sections of the paper are organized as follows: Section \ref{sec:bg} reviews the background on SNN compression and acceleration. Section \ref{sec:algorithm} details the algorithmic optimization framework proposed in FireFly-S. Section \ref{sec:strategy} describes the hardware acceleration strategy, encompassing both dataflow and sparse representation. Section \ref{sec:architecture} presents the hardware micro architecture of FireFly-S, featuring a sparse processor and dataflow orchestrator. Section \ref{sec:experiment} provides experimental results and evaluations. Section \ref{sec:conclusion} concludes the paper and discusses future work.

\section{Background} \label{sec:bg} 
\subsection{Spiking Neuron Model and Dynamic Equation}
Spiking Neural Networks utilize spiking neurons as their fundamental units, which communicate via binary spikes transmitted across weighted synapses. Although detailed models like the Izhikevich \cite{izhikevich2004model} and Hodgkin-Huxley \cite{hodgkin1952quantitative} neurons accurately capture biological neuron behavior, the computationally less demanding Integrate-and-Fire (IF) \cite{abbott1999lapicque} and Leaky Integrate-and-Fire (LIF) \cite{dayan2005theoretical} models are more commonly employed in current SNN implementations \cite{liu2022fpga} \cite{chen2022cerebron}.

The LIF neuron model simulates the passive decay of the membrane potential towards a resting state due to ion leakage, as well as the generation of a spike when the integrated input current drives the membrane potential above a certain threshold. This behavior is mathematically described by the following differential equation:
\begin{equation}
  \tau_m \frac{dV}{dt} = -V + R_m I(t)
\end{equation}
where $V$ is the membrane potential, $R_m$ is the membrane resistance, $\tau_m$ is the membrane time constant, and $I(t)$ is the input current. When the membrane potential reaches a threshold $V_{th}$, the neuron emits a spike and resets its potential to a resting value $V_{reset}$. As a discrete event-driven model, the LIF neuron can be described by the following equation:

\begin{equation}
  I_i(t) = \sum_{j} w_{ij} s_j(t) + b_i \label{eq:current_accumulation}
\end{equation}
  
\begin{equation}
  V_i(t) = V_i(t-1) + \frac{1}{\tau_m} \left( R_m I_i(t) - V_i(t-1) \right) \label{eq:potential_update}
\end{equation}
  
\begin{equation}
  V_i(t) = \begin{cases}
  V_{reset}, & \text{if } V_i(t) > V_{th} \\
  V_i(t), & \text{otherwise}
  \end{cases} \label{eq:spike_emission}
\end{equation}

Equation \eqref{eq:current_accumulation} accumulates the input current $I_i(t)$ from all presynaptic neurons $j$ with weights $w_{ij}$ and spikes $s_j(t)$, as well as a bias term $b_i$. Equation \eqref{eq:potential_update} updates the membrane potential $V_i(t)$ based on the input current and the membrane time constant. As for IF neurons, this decay process is absent, and the membrane potential simply integrates the input current without any leakage. Equation \eqref{eq:spike_emission} checks if the membrane potential exceeds the threshold $V_{th}$, emits a spike if true and resets the potential to $V_{reset}$. In practical implementations, the membrane time constant $\tau_m$ is often set to 2, and the potential resetting value $V_{reset}$ is set to 0, with the membrane resistance $R_m$ normalized to 1.

These spiking neuron models bring two key distinctions from traditional ANNs. First, the temporal dimension $T$ in SNNs involves the processing of membrane potentials over time, introducing temporal dependencies. Second, since neurons only fire spikes when their membrane potential exceeds a threshold, the neuron activity can be sparse \cite{yin2024workload} \cite{yin2024mint}. This binary and sparse nature of spike events presents both challenges and opportunities for efficient hardware implementation.

\subsection{Spiking Neural Networks Compression}
As previously mentioned, SNNs are inherently sparse and event-driven, making them well-suited for efficient processing on neuromorphic chips. However, hardware deployment is challenging due to increased computational complexity from temporal dynamics and a larger memory footprint from storing membrane potentials. To address these issues, compression techniques such as pruning and quantization have been explored to reduce the memory usage, and computational overhead, further enhancing the advantages of SNNs.

Accordingly, compressing SNNs through sparsity regularization and ADMM optimization was proposed by Deng \textit{et al.} \cite{deng2021comprehensive}. Chen \textit{et al.} \cite{chen2021pruning} introduced a gradient-based rewiring method, jointly optimizing weight values and connections for pruning. These pruning techniques provide two significant benefits for hardware implementation. 1) Lower memory footprint. The reduction of synaptic connections can lead to fewer spikes and inactive neurons, which decreases memory access and weight storage. Additionally, the reduced input current allows for a smaller bitwidth for membrane potentials. 2) Enhanced computational efficiency. Fewer operations are needed for pruned connections, reducing complexity in proportion to the sparsity. For a network with 75\% sparsity, the dense method processes 75\% of computations that are ineffectual on average. However, by incorporating sparsity detection logic, the same performance can be achieved with just a quarter of the processing elements, eliminating wasted computations.

For quantization, Putra \textit{et al.} \cite{putra2021q} developed the Q-SpiNN framework, which quantizes SNNs by exploring various quantization schemes and precision levels, guided by a reward function. Wei \textit{et al.} \cite{wei2024q} propose a Weight-Spike Dual Regulation method to mitigate the performance degradation caused by quantization. By reducing the bitwidth, quantization not only lowers the memory footprint but also directly addresses the computational overhead challenges in neuromorphic hardware by simplifying the arithmetic operations.

However, while pruning can induce dual-side sparsity in both weights and activations, and quantization can further boost the compression ratio, there is limited work on effectively combining these two techniques during training, especially in a hardware-friendly manner that considers the efficiency constraints presented by neuromorphic hardware for SNNs, such as fully integer arithmetic within neuron computations.

\subsection{Spiking Neural Networks Accelerators}
Hardware accelerators for SNNs can be broadly categorized into two approaches: general-purpose optimization and specialized acceleration. General-purpose methods focus on optimizing hardware architecture or modules for broader applicability. In contrast, specialized accelerators exploit the unique characteristics of SNNs, such as sparsity and event-driven dynamics, to achieve higher efficiency. For instance, Li \textit{et al.} presented FireFly \cite{li2023firefly} and FireFly v2 \cite{li2024firefly}, which leverage the DSP48E2 units in Xilinx FPGAs and a spatial-temporal dataflow to optimize the accelerator for overlay architectures. Targeting event-driven characteristics, Panchapakesan \textit{et al.} proposed SyncNN \cite{panchapakesan2022syncnn}, which encodes only activated neurons and quantizes LeNet to 4 bits. Chen \textit{et al.} introduced SiBrain \cite{chen2024sibrain}, featuring a sparse detection and response unit based on a channel-cache and block-multiplex design.

Despite these advancements, current architectures often fail to utilize sparsity effectively, focusing primarily on spike sparsity and neglecting dual-side sparsity. Furthermore, many designs rely on overlay architectures that necessitate off-chip memory access, which can be inefficient. Notably, networks are typically quantized to 8 to 16 bits \cite{li2023firefly} \cite{li2024firefly} \cite{chen2024sibrain}, leading to increased storage demands. In contrast, SyncNN's 4-bit quantization supports only a lightweight LeNet model \cite{panchapakesan2022syncnn}, highlighting limitations in scalability. There remains a significant gap in research on fully on-chip spatial architectures that could enhance throughput and power efficiency by facilitating entire inference processes on-chip. This underscores the urgent need for software-hardware co-design methodologies that can exploit both dual-side sparsity and the potential of low-bit quantization in SNNs, thereby enabling more efficient computation and reduced memory footprints.

\section{Algorithmic Optimization} \label{sec:algorithm}
As discussed earlier, while SNNs offer inherent spike sparsity, they also pose challenges in terms of computational complexity and memory footprints. This section first discusses compression sequence, focusing on the order of pruning and quantization. We then introduce how FireFly-S implements pruning and quantization during training, where pruning enables hardware exploitation through dual-side sparsity and quantization further enhances the efficiency of computation and memory.

\subsection{Compression Sequence}
The sequence of pruning and quantization shows impact on the performance of neural networks. Some studies advocate for a quantize-then-prune approach \cite{zhao2019focused} \cite{yu2020joint}, while others, such as Zhang \textit{et al.} \cite{zhang2021training}, suggest that the optimal sequence may depend on the network architecture. This indicates that different networks exhibit varying tendencies regarding the order of compression techniques. Therefore, treating pruning and quantization as separate steps may lead to suboptimal performance across various networks, and any necessary post-processing or recalibration \cite{lazarevich2021post} \cite{shang2024enhancing} to maintain accuracy can become complex and challenging.

\begin{figure*}
  \centering
  \includegraphics[width=1.0\linewidth]{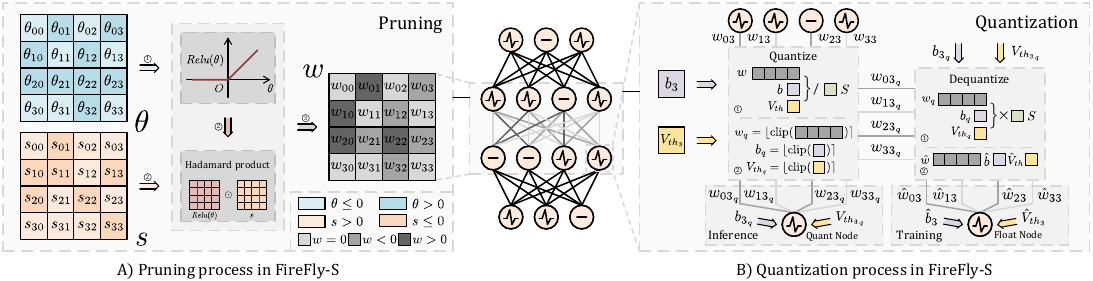}
  \caption{
    Compression process in FireFly-S. A) Pruning process. B) Quantization process, including both inference and training stages.
  }
  \label{fig:compress}
\end{figure*}

An alternative strategy is to jointly optimize pruning and quantization, where weight values and connections are simultaneously refined. This integrated approach explores a broader search space and has the potential to yield more optimal solutions \cite{park2022quantized} \cite{zandonati2023towards}. By embedding pruning and quantization within the training process, this strategy not only enhances the balance between compression and accuracy but also eliminates the need for additional post-training steps, such as retraining or fine-tuning. This joint optimization is precisely the strategy employed by FireFly-S, as depicted in Fig. \ref{fig:compress}.

\subsection{Compression Methodologies}
The pruning algorithm we adopt is based on gradient rewiring \cite{chen2021pruning}. In this method, each synaptic weight $w$ is represented as $w = s \odot \text{ReLU}(\theta)$, where $\theta$ represents the synaptic parameter, and $\odot$ denotes element-wise product. The sign factor $s$, determined by the sign of the initial weight, indicates whether the connection is excitatory or inhibitory. This representation is illustrated in the left part of Fig. \ref{fig:compress}. During training, $\theta$ is influenced by a Laplace prior distribution to impose a regularization effect. Specifically, a penalty term $-\alpha \, \text{sign}(\theta - \mu)$ is applied, where $\alpha$ and $\mu$ represent the scale and location parameters, respectively\cite{chen2021pruning}. This penalty pushes some $\theta$ values towards non-positive numbers, thereby eliminating the corresponding connections.

Our approach builds on this by applying weight decay to $\theta$, scaling $\theta$ by $(1 - \eta \lambda)$ during each update, where $\eta$ is the learning rate and $\lambda$ as the weight decay factor. This introduces an $\ell_2$ regularization effect, which not only encourages more $\theta$ values to approach zero, promoting sparsity, but also helps avoid sharper minima and improves generalization during optimization \cite{wilson2017marginal} \cite{zhou2020towards}. Besides, to enable more comprehensive network compression, we apply this pruning strategy to both convolutional and fully connected layers, including bias parameters that are optimized by AdamW \cite{loshchilov2017decoupled}.

Due to the adoption of fine-grained pruning, there is a likelihood that some weights within an output channel dimension might be entirely zero. In ANNs, even if the weights are zero, the presence of a bias can still result in a non-zero output for a layer. Thus, the kernel of that output channel cannot be pruned solely based on weight values. However, SNNs exhibit different behavior due to the presence of a neuron threshold. In SNNs, even the accumulated bias across the temporal dimension might not exceed the threshold, particularly when the neuron model includes leaky currents. Consequently, if the neuron remains inactive (silent) throughout, it indicates that the kernel of that output channel could potentially be pruned. Algorithm \ref{algo:pruning} provided simulates the inference process to identify such silent output channels and prunes them accordingly.

\begin{algorithm}
  \label{algo:pruning}
  \caption{Temporal Membrane Potential Evaluation for Channel Pruning}
  \KwIn{
    \begin{itemize}
      \item Number of output channels $C_o$
      \item Number of time steps $T$
      \item Array of thresholds $\mathbf{V}_{\text{th}}$ (one threshold per channel)
      \item Array of biases $\mathbf{b}$ (one bias per channel)
    \end{itemize}
  }
  \KwOut{List of active channels $A$}
  
  Initialize membrane potentials matrix $V$ of size $C_o \times T$ to zero\;
  Initialize list of active channels $A \leftarrow []$\;
  
  \For{$c_o \leftarrow 0$ \KwTo $C_o-1$}{
    \For{$t \leftarrow 0$ \KwTo $T-1$}{
      $V[c_o][t] \gets V[c_o][t-1] + \mathbf{b}[c_o]$\;
      \If{$V[c_o][t] > \mathbf{V}_{\text{th}}[c_o]$}{
        $A \leftarrow A \cup \{c_o\}$\;
        \textbf{break}\;
      }
    }
  }
  
  \Return{$A$}\;
\end{algorithm}

For quantization, we adopt the principles of Learned Step Size Quantization (LSQ) \cite{esser2019learned}. To enable gradient-based optimization of the quantization scale, LSQ employs the straight-through estimator (STE) \cite{bengio2013estimating}, which approximates the gradient of the otherwise non-differentiable quantization operation. The quantization process is defined as:
\begin{align}
  r_q &= Q(r, S, Q_N, Q_P) \\
    &= \lfloor\operatorname{clip}(r/S, -Q_N, Q_P)\rceil \\
  \hat{r} &= r_q \times S 
\end{align}

Here, $r$ represents the real-valued input, $S$ the quantization scale, $r_q$ the quantized value, and $\hat{r}$ the dequantized value. The function $Q(\cdot)$ denotes the quantization operation and $\lfloor \cdot \rceil$ denotes rounding to the nearest integer. $Q_N$ and $Q_P$ define of the minimum and maximum quantized absolute values, respectively, ensuring the clipped value remains within the designated range. The gradient of the quantization error with respect to the scale $S$ is:
\begin{equation}
  \frac{\partial \hat{r}}{\partial S} = 
  \begin{cases} 
  -r/S + \lfloor r/S \rceil & \text{if } -Q_N < r/S < Q_P \\
  -Q_N & \text{if } r/S \leq -Q_N \\
  Q_P & \text{if } r/S \geq Q_P 
  \end{cases}
\end{equation}

While LSQ has proven effective for quantizing ANNs, it is not directly applicable to SNNs and is not optimized for fully integer computation. This is because the bias of layers and the threshold of spiking neurons are still real-valued, and using the same quantization scale optimized for weights on the bias and threshold can lead to accuracy degradation. 

Building on the success of STE in approximating gradients for quantization, we introduce a novel LSQ variant tailored specifically for SNNs. The right part of Fig. \ref{fig:compress} illustrates the quantization process in FireFly-S. During training, weights, biases, and thresholds are quantized and dequantized, with the dequantized values used for forward and backward propagation. In contrast, inference uses only quantized parameters for forward propagation, removing the need for dequantization, scale storage, and related computations.

Our approach differs from existing methods in two key ways: 1) We employ learnable per-channel scales, rather than relying on statistical methods like min-max or percentile-based approaches, which are commonly used in frameworks like SyncNN and earlier versions of the FireFly series. 2) Unlike standard LSQ and quantization frameworks like TensorRT, which often neglect bias quantization, our method jointly optimizes shared quantization scales for weights, biases, and neuronal thresholds within each channel during training.

These two key distinctions also offer two major benefits: 1) Learnable per-channel scales enable the model to capture and address quantization errors arising from various computations during training, including neuron dynamics like the decay process. This results in improved accuracy, as demonstrated in Section \ref{sec:experiment}. This end-to-end training process ensures that, once trained, the model can be deployed directly without any additional precision loss. In contrast, statistical methods require a post-training process, and since the scales are based on weight statistics, applying them directly to biases and neuronal thresholds can cause mismatches and lead to accuracy degradation. 2) By sharing the quantization scale across weights within the same channel, along with the corresponding bias and neuronal threshold, our approach ensures consistent scaling across parameters. This eliminates the need for rescaling during weight accumulation, bias addition, and threshold comparison, enabling efficient integer-only inference. Additionally, due to the unique properties of spiking neurons, there is no need to rescale per-channel activations during inference compared to ANN quantization, as the per-channel quantization scales are absorbed into the membrane thresholds. Based on these characteristics, our hardware accelerator can perform all computations using integer arithmetic, simplifying the computational logic and memory footprint.

By jointly optimizing quantization and incorporating pruning, while explicitly considering the unique characteristics of SNNs, this approach enables high compression rate and fully integer computation deployment for more efficient inference. Section \ref{sec:experiment} provides a detailed analysis of accuracy and compression outcomes across various networks, benchmarked against E3NE \cite{gerlinghoff2021e3ne} and SyncNN \cite{panchapakesan2022syncnn}.

\section{Hardware Acceleration Strategy} \label{sec:strategy}
To unlock the full potential of inter-layer pipelining in hardware acceleration, a carefully chosen parallelism scheme and a consistent dataflow pattern are essential. Besides, efficiently exploiting the dual-side sparsity of SNNs hinges on a suitable sparse representation that minimizes memory footprint while maximizing throughput. This section delves into the specific dataflow and parallelism strategies employed by FireFly-S, followed by an analysis of the chosen sparse representation and its impact on hardware acceleration efficiency. 

\subsection{Dataflow and Parallelism Schemes}
FireFly v2 demonstrated the importance of carefully designed dataflow in SNN implementations, focusing on an overlay architecture to balance spatial and temporal dimensions. FireFly-S, however, takes a different route by embracing a spatial architecture, an approach whose dataflow considerations remain unexplored for SNNs on FPGAs. The challenge lies in fully harnessing the potential of spatial architectures, which offer complete network pipelining but demand meticulously designed dataflow to ensure consistent and efficient data movement between input and output.

\begin{figure*}
  \centering
  \includegraphics[width=1.0\linewidth]{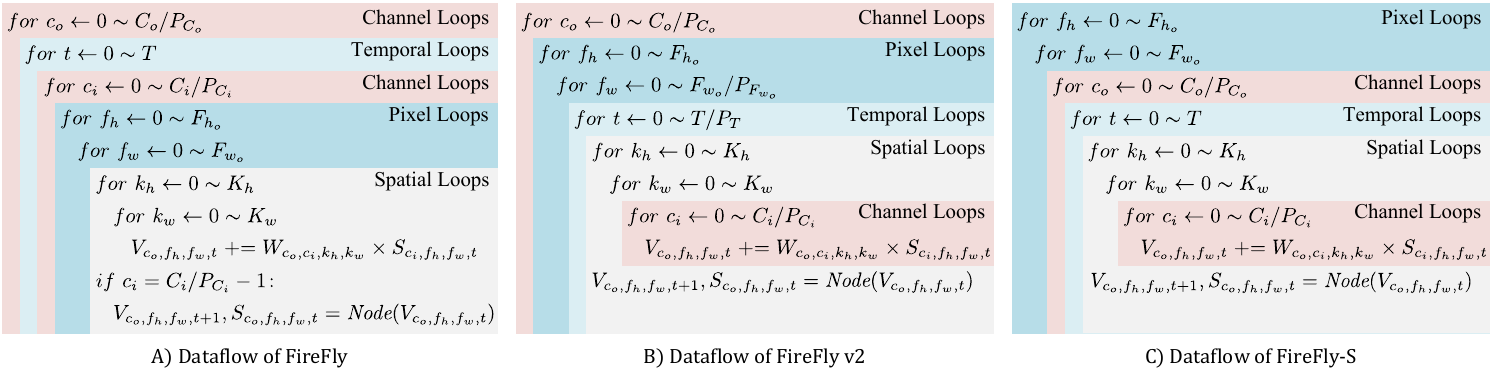}
  \caption{
  Comparison of the dataflow in FireFly-S to its previous version FireFly and FireFly v2. A) FireFly's dataflow design. B) FireFly v2's dataflow design. C) FireFly-S's dataflow design.
  }
  \label{fig:dataflow}
\end{figure*}

Fig. \ref{fig:dataflow} illustrates the dataflow evolution within the FireFly series. Here, $C_o$, $C_i$, $F_{h_o}$, $F_{w_o}$, $K_h$, $K_w$, and $T$ represent the number of output channels, input channels, output feature map height, output feature map width, kernel height, kernel width, and temporal dimension, respectively. $V$, $W$, and $S$ denote membrane voltage, synaptic weight, and activation spike. Parallelism levels are denoted by $P$, with $P_{C_o}$, $P_{C_i}$, $P_{F_{w_o}}$, and $P_T$ corresponding to parallelism across output channels, input channels, output feature map width, and the temporal dimension. 

FireFly's initial design, with its outer-loop temporal processing, suffers from inefficiencies compared to the spatial-temporal approach of FireFly v2. This is because handling the temporal dimension in the outer loop necessitates storing membrane voltage values for extended periods, resulting in a significant memory footprint proportional to $P_{C_o} \times F_{h_o} \times F_{w_o}$. FireFly v2 mitigates this by processing the temporal dimension in a more inner loop, effectively minimizing the impact of long-term temporal dependencies. This optimization reduces the required memory storage for membrane voltage to $P_{C_o} \times P_{F_{w_o}}$, a value determined by the level of parallelism. 

However, a key distinction arises between FireFly-S and its predecessors. While both FireFly and FireFly v2 utilize overlay architectures, relying on off-chip memory for feature map and weight storage, FireFly-S adopts a spatial architecture to enable full inter-layer pipelining. This shift necessitates a departure from FireFly v2's dataflow strategy. In detail, specific solutions are required to address the two main challenges posed by spatial architectures: 1) the tight on-chip memory constraints, which demand efficient memory utilization, and 2) the need for a consistent dataflow pattern to ensure efficient processing across layers and avoid pipeline stalls.

Directly applying FireFly v2's dataflow to a spatial architecture would prove inefficient. FireFly v2, designed with off-chip memory reliance, employs an output-channel outermost loop. This approach, while functional, leads to excessive on-chip memory consumption in a spatial architecture. Computing each output channel requires access to all input feature maps, necessitating storage of all layer feature maps on-chip, amounting to 
$L \times F_{h_o} \times F_{w_o} \times T \times C_i$ memory, where $L$ represents the number of layers. Such demands would exceed the limits of on-chip memory.

Besides, the output channel in this layer is the input channel in the next layer, so it is challenging to design a consistent dataflow pattern. But we can minimize the storage and data movement. Employing an output-channel outermost loop and an input-channel innermost loop, as in FireFly v2, would lead to significant pipeline stalls in a spatial architecture. Each layer would need to wait for all output channel computations of the previous layer to complete, consequently, it could not match the need of non-stall pipeline for spatial architecture.

\begin{table}[!t]
  \renewcommand\arraystretch{1.25}
  \caption{Memory Storage Comparison of FireFly-S\\ with Previous Versions\label{tab:dataflow}}
  \centering
  \begin{tabular}{c|c|c}
  \hline
   & $V_{buf}$ & $S_{buf}$ \\
  \hline
  \hline
  FireFly & $P_{C_o} \!\times\! F_{h_o} \!\times\! F_{w_o}$ & $F_{h_o} \!\times\! F_{w_o} \!\times\! T \!\times\! C_i$ \\
  \hline
  FireFly v2 & $P_{C_o} \!\times\! P_{F_{w_o}}$ & $F_{h_o} \!\times\! F_{w_o} \!\times\! T \!\times\! C_i$ \\
  \hline
  FireFly-S & $P_{C_o}$ & $((K_h \!-\! 1) \!\times\! F_{w_o} \!+\! K_w) \!\times\! T \!\times\! C_i $ \\
  \hline
  \end{tabular}
\end{table}

To address these challenges, we have restructured the dataflow by placing the channel loops inside the pixel loops while maintaining the temporal loop nested within the output channel loops. This configuration provides two main benefits. Firstly, by confining the temporal loop, it allows the membrane voltage to accumulate consecutively, removing the need to store values for each time step. Secondly, by embedding the channel loop within the pixel loop, it eliminates long-term dependencies between channels, thereby minimizing data movement and ensuring a consistent dataflow. This optimization retains only the essential feature map data necessary for network operations and data reuse, effectively reducing the required feature map storage by a factor of $F_{h_o} / (K_h - 1)$. Table \ref{tab:dataflow} presents the results of this optimization, detailing the membrane voltage and feature map storage requirements, denoted by $V_{buf}$ and $S_{buf}$ respectively.

\subsection{Sparse Representation Analysis} \label{subsec:sparse}
While pruning and the inherent spike sparsity in SNNs yield high dual-side sparsity, the choice of sparse representation significantly impacts the memory footprint and computational efficiency of hardware accelerators. Several sparse representation schemes exist, including Coordinate Lists (COO), Compressed Sparse Row (CSR), and Bitmap representations. 

Coordinate Lists (COO) offer a simple and intuitive representation, storing non-zero values alongside their corresponding indices. Compressed Sparse Row (CSR) improves upon this by compressing row indices, storing row offsets instead of explicit indices. While requiring additional decoding logic to locate non-zero values, CSR reduces memory footprint by eliminating redundant row index storage. Bitmap representation takes a different approach, storing non-zero values with a corresponding mask indicating their presence. This method implicitly encodes addresses within the mask, eliminating the need for explicit indices. 

This difference in encoding strategies leads to a key trade-off: While COO and CSR formats offer the advantage of storing only the indices of non-zero values, these indices can require a larger number of bits compared to the single bit used in Bitmap representation. Bitmap, on the other hand, utilizes a bit array to represent the presence or absence of all values, regardless of whether they are zero or non-zero. This difference in storage strategy means that the sparsity of the matrix plays a crucial role in determining which format is more memory-efficient. 

Furthermore, decoding non-zero pairs in a dual-side sparse matrix using COO and CSR necessitates separate decoding for weights and spikes. This separate decoding can lead to pipeline stalls when mismatches occur, ultimately reducing overall throughput. Bitmap representation, however, allows for simultaneous decoding and direct identification of matching non-zero pairs, mitigating these performance bottlenecks.

Therefore, considering the critical importance of both memory storage and throughput in hardware design, we conduct an experimental analysis to evaluate these trade-offs. Throughput is measured as the number of non-zero pairs processed per cycle, while storage is determined by calculating the number of bits required to store the sparse matrix in each format. Both CSR and COO decoding logic are assumed to decode one non-zero value per cycle for both weight and spike decoders concurrently, and Bitmap decoding logic is assumed to decode one non-zero pair per cycle for each vector pair with a length of 16/32/64/128, which indicates the parallelism for a layer in our hardware architecture. Notably, the Bitmap logic still consumes one clock cycle when there is no non-zero pair for the current vector pair.

\begin{figure}
  \includegraphics[width=1\linewidth]{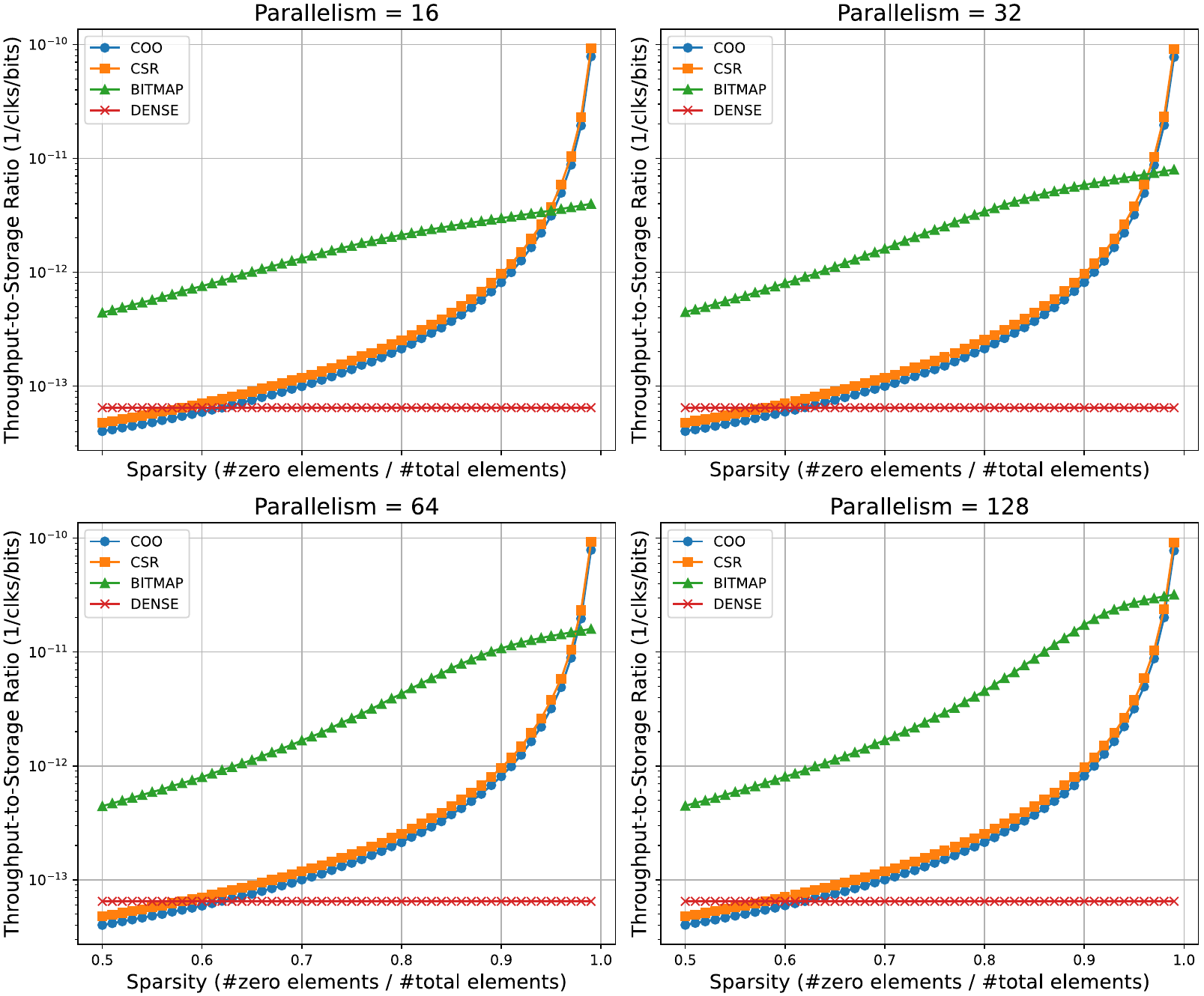}
  \caption{
  Throughput-to-storage ratio versus sparsity for COO, CSR, and Bitmap sparse representations and dense representation. Data points are plotted at 0.01 sparsity intervals.
  }
  \label{fig:Parallel163264128}
\end{figure}

This analysis, conducted within a specific experimental environment, simulating a convolutional operation with parameters $(T, C_o, C_i, F_{h_o}, F_{w_o}, K_h, K_w) = (4, 64, 32, 16, 16, 3, 3)$. The setup represents a single layer within the 9-layer network for CIFAR-10 recognition discussed in Section \ref{sec:experiment}, and the selection does not affect the overall observed trend. Fig. \ref{fig:Parallel163264128} illustrates the throughput-to-storage ratio across various sparsity levels, comparing the performance of COO, CSR, and Bitmap sparse representations against a dense representation. While COO and CSR outperform Bitmap at extremely high sparsity levels (above 95\%) due to Bitmap's larger memory footprint, Bitmap consistently demonstrates superior performance at other sparsity levels. This advantage is further amplified with larger vector pair lengths, extending its applicability to a wider range of sparsity. Notably, in Figure \ref{fig:Parallel163264128}, the sparsity level is shared by both synaptic weights and activation spikes. Experiments indicate that the sparsity of activation spikes ranges from 70\% to 90\%. Therefore, the sparsity of the weights must be significantly higher to reach the threshold where Bitmap becomes less efficient than COO or CSR. Given these observations, we adopt the Bitmap representation in FireFly-S, prioritizing its greater generality and higher throughput-to-storage ratio across a wide range of sparsity levels.

\section{Hardware Micro Architecture} \label{sec:architecture}
\begin{figure}
  \includegraphics[width=1\linewidth]{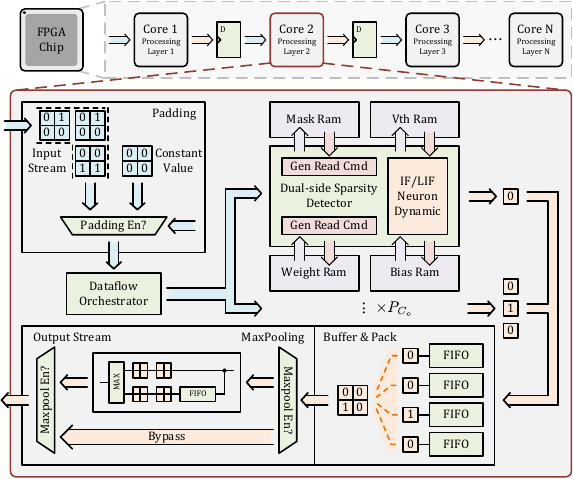}
  \caption{
    Inter-layer Pipelined Architecture of FireFly-S. All cores are mapped onto a single FPGA chip, with each core acting as a pipeline stage responsible for the inference of one layer. These stages integrate modules for Padding, Dataflow Orchestration, Dual-side Sparsity Detection, and Maxpooling.
  }
  \label{fig:architecture}
\end{figure}

The hardware microarchitecture of FireFly-S is intricately designed to harness the dual-side sparsity characteristic of SNNs. It features a spatial layout that incorporates inter-layer pipelining, as illustrated in Fig. \ref{fig:architecture}. Each pipeline stage is specifically designed to handle a single layer of inference within the FPGA chip, equipped with on-chip parameters and specialized modules including Padding, Dataflow Orchestrator, Dual-side Sparsity Detector, and Maxpooling.

The Padding module dynamically adjusts input dimensions for convolutional layers, adding padding in "same" mode to match output and input sizes, while bypassing padding for "valid" mode or fully connected layers. The Dataflow Orchestrator is pivotal in ensuring consistent dataflow between layers to maintain pipeline efficiency. It also efficiently manages the im2col operation for SNNs, essential for reorganizing input data into a format suitable for convolution operations, all while maintaining a membrane potential storage-free approach. The Dual-side Sparsity Detector module optimizes computations by identifying matching non-zero elements in both activation spikes and synaptic weights, enabling computational savings by bypassing unnecessary operations. Then, it generates specific read commands for relevant data (weights, biases, etc.) and efficiently updates neuron dynamics in the IF/LIF neuron model. Finally, the optional Maxpooling module provides spatial downsampling of the output, the output stream is then pipelined and fed into the next layer for further processing.

This section provides an in-depth exploration of the key elements of FireFly-S, particularly the Dataflow Orchestrator and the Dual-side Sparsity Detector, detailing their essential roles and functionalities within the broader system architecture.

\subsection{Dataflow Orchestrator}
As discussed in Section \ref{sec:strategy}, spatial architectures require a consistent dataflow pattern. However, achieving this consistency presents a challenge due to the conflicting dataflow requirements between the output of one layer and the input of the next layer. Specifically, the membrane potential storage-free calculation in FireFly-S necessitates placing the output channel loop \textbf{outside} the temporal loop. This arrangement contrasts with the input dataflow of the subsequent layer, which requires the channel loop to be \textbf{inside} the temporal loop for efficient processing. Furthermore, convolution operations introduce an additional complexity: the need for the im2col operation to transform the input feature map into a matrix format suitable for convolution.

\begin{figure*}
  \centering
  \includegraphics[width=1.0\linewidth]{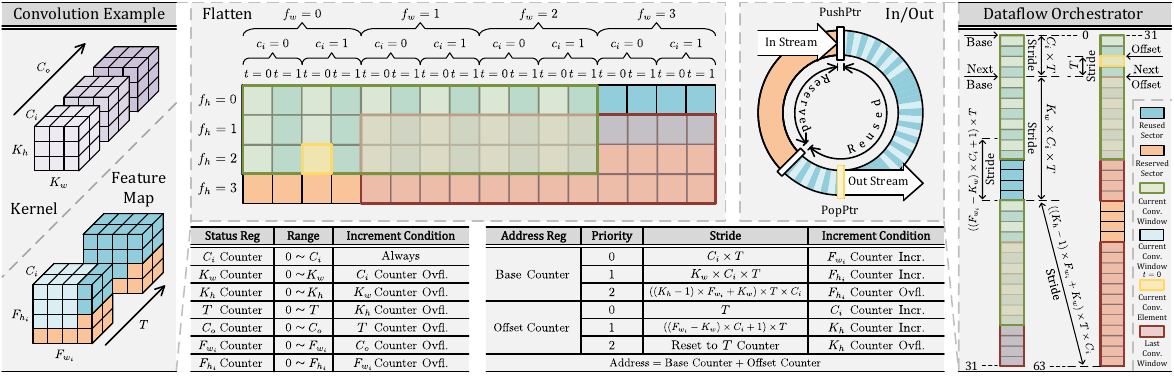}
  \caption{
    Dataflow Orchestrator in FireFly-S, illustrated with an example where $(T, C_o, C_i, F_{w_i}, F_{h_i}, K_{w_i}, K_{h_i}) = (2,\ 2,\ 3,\ 4,\ 4,\ 3,\ 3)$ and $(P_{C_o}, P_{C_i}) = (1,\ 1)$. The "Priority" column in the table indicates the order in which stride is taken, with higher numbers representing higher priority.
  }
  \label{fig:orchestrator}
\end{figure*}

To reconcile these conflicting dataflow requirements and manage the im2col operation, FireFly-S employs a dedicated Dataflow Orchestrator, as illustrated in Fig. \ref{fig:orchestrator}. This FIFO interface logic utilizes on-chip RAM divided into two sectors. The first sector is essential for three key functions: 1) dataflow permutation across the temporal and channel dimensions, 2) execution of the implicit im2col operation, and 3) data reuse for continuous operation when parallelism cannot fully unroll the convolution. The size of this sector is determined by the input dimensions and kernel size, allowing it to store only the essential data needed to maintain the convolution window while the input feature map streams in. Additionally, both dataflow permutation and data reuse occur within the convolution window, with the reuse time calculated as $C_o / P_{C_o}$. The remaining on-chip RAM serves as a configurable holding area, acting as a buffer for incoming spike feature maps from upstream processing. This holding area prevents pipeline stalls that would otherwise occur when the reused sector is full. Additionally, it provides a transition buffer for switching convolution windows. For clarity, the example in Fig. \ref{fig:orchestrator} depicts a parallelism level of 1, with an input dataflow of $(F_{h_i}, F_{w_i}, C_i, T)$ being transformed to the output dataflow of $(F_{h_i}, F_{w_i}, C_o, T, K_h, K_w, C_i)$.

The orchestrator leverages a combination of status registers, implemented as simple counters, and some stride counters. The status registers, chained to form loop iterators, pinpoint the current position within the dataflow. These registers don't directly determine memory addresses but provide signals that guide the stride counters in generating the correct addresses for fetching data from the RAM. The use of stride counters for base address and offset address, each with its own specific increment value, allows the orchestrator to navigate the non-sequential memory access patterns required by the implicit im2col, data transpositions and data reuse, eliminating the need for computationally expensive multiplication operations.

The orchestrator manages data access using a push pointer and a pop pointer, reflecting the principles of a traditional FIFO buffer. While the logic for the push pointer aligns with standard FIFO operations, the pop pointer is determined by the combined values of the base and offset counters. The base counter indicates the starting position of the current convolution window within the RAM, while the offset counter navigates through the elements within that window. 

For \textit{valid} convolution with stride 1, the base counter increment for the dataflow coordinator adapts based on the window's position within the feature map. Under normal operation, the increment is simply $C_i \times T$, reflecting the number of data elements in the previous window's channel and temporal dimensions. In addition to this regular increment, specific adjustments are needed for efficient transitions between rows and feature maps. After processing the last window of a row, the base counter must jump to the beginning of the next row, requiring an increment of $(K_w \times C_i \times T)$ to bypass the remaining $(K_w - 1)$ columns of the previous row. Similarly, transitioning to a new feature map necessitates an increment of $((K_h - 1) \times F_{w_i} + K_w) \times C_i \times T$ to address the start of the new feature map's data. This ability to directly access data in RAM, regardless of position, boosts the efficiency of the Dataflow Orchestrator. In contrast, traditional line buffers, which rely on sequential shifting, introduce overhead during row and feature map transitions, as they must shift data sequentially to align with the required starting position for processing.

The offset counter works in conjunction with the base counter to pinpoint the exact data address within each convolution window. To prioritize the channel loop over the temporal loop for efficient processing, the offset counter typically increments by $T$, reordering the data stream from the input's temporal-within-channel organization to a channel-within-temporal organization. When transitioning between rows within the convolution window, the offset counter's stride changes to $((F_{w_i} - K_w) \times C_i + 1) \times T$ to account for the memory layout shift between rows. Furthermore, during membrane potential storage-free calculations, the offset counter dynamically adapts when switching iterators in the temporal dimension. It resets to the value stored in the temporal counter within the status register and then continues incrementing using the previously described stride patterns. 

After processing the convolution window, the offset counter resets to its initial value, while the base counter remains unchanged during data reuse. Once the reuse is complete, the base counter increments to indicate the starting position of the next convolution window, as previously depicted. This cyclical process ensures that all output channels $C_o$ are effectively traversed while reusing the input data within the constraints of limited $P_{C_o}$ parallelism.

Besides its role in managing the dataflow, the design logic of Dataflow Orchestrator also leverages the regularity of data access patterns, which are dictated by consistent stride patterns, to pre-calculate base and offset counter values. This pre-calculation enables pipelined address generation for the pop pointer, efficiently determining the next data element for processing. The predictable nature of these strides also facilitates the implementation of the Dataflow Orchestrator's full and empty signals as registers, reducing combinational logic delays and enhancing timing closure. Furthermore, as data dimensions increase, the bit width of the status and address registers scales logarithmically. This scaling ensures that the logic remains efficient and does not become a bottleneck during network expansion. Ultimately, these design choices support a working frequency of 333~MHz across our entire architecture, as detailed in Section \ref{sec:experiment}. 

\begin{figure*}
  \centering
  \includegraphics[width=1.0\linewidth]{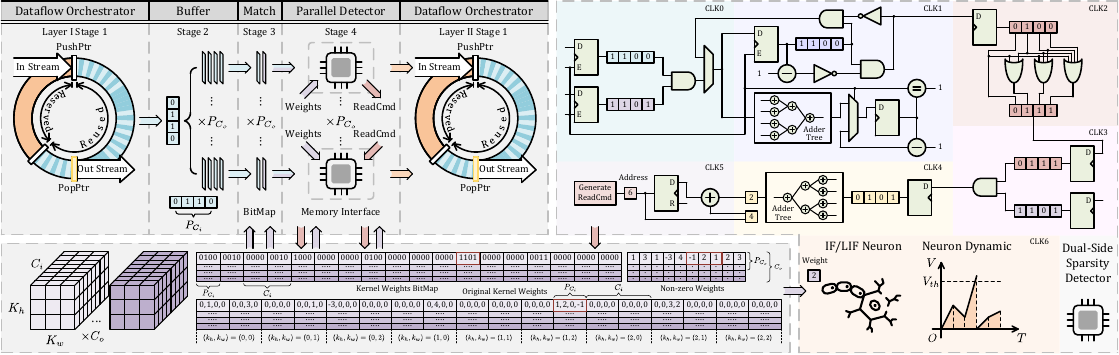}
  \caption{
    Schematic of the Dual-side Sparsity Detector in FireFly-S, illustrated with configuration parameters $(C_o, C_i, K_{w_i}, K_{h_i}) = (4,\ 8,\ 3,\ 3)$ and parallelism parameters $(P_{C_o}, P_{C_i}) = (2,\ 4)$.
  }
  \label{fig:detector}
\end{figure*}

\subsection{Dual-side Sparsity Detector}
After passing through the Dataflow Orchestrator, the activation spikes are buffered and aligned with their corresponding weight mask data, which is represented in Bitmap format. This prepared data is then processed by the Dual-side Sparsity Detector, as illustrated in Fig. \ref{fig:detector}.

The lower left corner of Fig. \ref{fig:detector} depicts the initial transformation of the original sparse weight matrix into Bitmap format. This transformation, conducted offline before hardware deployment, evaluates each matrix element to determine its non-zero status, resulting in the creation of a bitmap mask alongside a separate collection of non-zero values. These bitmaps are then segmented into vectors, each sized according to the input channel parallelism, $P_{C_i}$. This segmentation, also performed pre-deployment based on workload estimations detailed in Section \ref{subsec:parallelism}, ensures balanced processing across layers to mitigate pipeline stalls. 

Following segmentation, both the bitmaps and non-zero values are distributed across the output channel parallelism, $P_{C_o}$. Consequently, each detector within the Dual-side Sparsity Detector is assigned a segment of the non-zero weight RAM along with its corresponding bitmap mask RAM, where the length of each bitmap segment corresponds to $P_{C_i}$. This arrangement ensures that each detector operates with a precise portion of data, exploiting the fact that computations for each output channel are independent and can be performed in parallel. A higher $P_{C_o}$ allows for more detectors to operate in parallel, further enhancing throughput. This higher $P_{C_o}$, representing the number of parallel output elements calculated in this layer, directly translates to a potentially higher $P_{C_i}$ in the next layer. This is because these parallel output elements are then assembled into a complete output vector, which in turn, forms the input spikes vector for the subsequent layer, as shown in the upper left corner of Fig. \ref{fig:detector}.

The processing flow within each detector, as depicted in the upper right corner of Fig. \ref{fig:detector}, begins with the reception of input spikes alongside a corresponding weight mask. At CLK0, a bitwise AND operation is executed between the vector pair, yielding a result denoted as $x$, which identifies active events. Following this, at CLK1, an adder tree efficiently tallies the number of non-zero pairs in $x$, and this count is stored in a decrement counter. The counter serves a dual purpose: it indicates the completion of processing non-zero pairs and signals the readiness to receive a new input vector pair.

In this stage, a one-hot encoded output, $y$, is generated through the logical expression $y = x \wedge \neg(x - 1)$. This operation positions a '1' at the index of the first non-zero pair in $x$. The register holding $x$ is then updated using the expression $x = x \wedge \neg y$, which ensures that processed pairs are not reactivated in subsequent cycles. This process repeats until all non-zero pairs in $x$ are processed, producing a series of one-hot encoded output vectors $y$ that accurately represent the activation pattern for the current input.

The next stage utilizes the one-hot vector $y$ to generate a new vector with bits set from the least significant bit up to the index where '1' is set in $y$. For example, if $y$ is 4'b0100, which indicates the third position, the generated vector would be 4'b0111. This vector is then subjected to a bitwise AND operation with the weight mask to pinpoint active indices that align with the mask. Continuing with the example, if the weight mask is 4'b1101, the result of the AND operation between 4'b0111 and 4'b1101 would be 4'b0101. This result, obtained at CLK3, allows the detector at the following cycle (CLK4) to determine the precise storage offset — 2 in this example — of the weight data corresponding to the '1' in the original one-hot vector 4'b0100 using an adder tree. At CLK5, the calculated offset is accumulated into an offset register. This accumulated offset is then added to the base address stored in a base address register to generate the final address for accessing the weight data, thus preparing the read command for the specific weight data. Finally, at CLK6, the weight data is fetched from the RAM and used to perform IF/LIF neuron model computation. It is important to note that the membrane potential is stored in a register, and due to the dedicated dataflow, the voltage for each time step is temporally stored in the register and reused for the next time step.

For clarity, Fig. \ref{fig:detector} does not explicitly illustrate the interactions with bias and threshold RAMs as depicted in Fig. \ref{fig:architecture}. However, these interactions are straightforward: each output channel's neuron dynamics require the addition of bias, and the membrane potential must be compared with the threshold to determine whether a spike should be generated. Since both bias and threshold are assigned per channel (reflecting per-channel quantization), and because the bias is always added and the threshold is always checked against the membrane potential to determine spike generation, the reading addresses for both are represented by an incrementing counter.

Moreover, to improve efficiency during the dual-side sparse acceleration process involving weight and spike vectors, a specific optimization is employed. Typically, if the bitwise AND operation between the spike vectors and the weight vector bitmap results in all zeros, it would lead to a wasted cycle due to the absence of valid non-zero pairs. To counter this inefficiency, we incorporate the bias, treated as a constant input of '1' with a corresponding weight equal to the bias value, which is processed in each time step. By designing logic to absorb the cycle that would have been wasted on detection into the cycle used for adding the bias, we ensure that the bias effectively replaces the ineffective cycle. This eliminates unnecessary idle clock cycles and enhances the efficiency of the Dual-side Sparsity Detector.

\begin{figure}
  \includegraphics[width=1\linewidth]{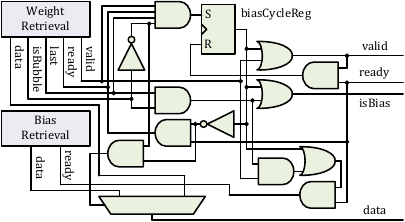}
  \caption{
    Optimized logic for weight and bias retrieval in the Dual-side Sparsity Detector, illustrating the strategic implementation of bias addition to mitigate idle detector cycles.
  }
  \label{fig:squeeze}
\end{figure}

Fig. \ref{fig:squeeze} depicts the optimized logic employed in the weight and bias retrieval process. This process utilizes handshake logic with \textit{valid} and \textit{ready} signals; however, since the bias is always considered valid, its corresponding \textit{valid} signal is disregarded. Central to this mechanism is the \textit{biasCycleReg}, a register determining the need to utilize bias addition to mitigate idle cycles. Specifically, when the weight retrieval logic's \textit{isBubble} signal is low and the \textit{last} signal is high—indicating an impending time step increment—the \textit{biasCycleReg} is set high, signaling that the current data pertains to weight (\textit{isBias} is low), and thus, the bias will be added in the subsequent cycle. Conversely, if \textit{isBubble} is high and \textit{last} is also high, the \textit{biasCycleReg} resets, allowing for the immediate addition of the bias within the current cycle to optimize efficiency by eliminating idle cycles.


\section{Implementation and Experiments} \label{sec:experiment}

\subsection{Experiments Setup}

\begin{table}[!t]
  \renewcommand\arraystretch{1.25}
  \caption{Configuration of SNN Models}
  \centering
  \begin{tabular}{c|c}
  \hline
  Model & Configuration \\
  \hline
  \hline
  \multirow{2}{*}{SCNN5} & 1x28x28-8c3p1-16c3p2-mp2-32c3p1- \\
  & mp2-64c3p1-64c3p1-mp2-10fc\tnote{3} \\
  \hline
  \multirow{2}{*}{SCNN7} & 2x48x48-16c3p1-32c3p1-mp2-32c3p1-64c3p1- \\
  & mp2-64c3p1-128c3p1-mp2-128c3p1-10 \\
  \hline
  \multirow{2}{*}{SCNN9} & 3x32x32-16c3p1-32c3p1-mp2-32c3p1-64c3p1-mp2- \\
  & 64c3p1-128c3p1-mp2-128c3p1-256c3p1-mp2-10 \\
  \hline
  \end{tabular}
  \label{tab:models}
\end{table}

To evaluate the performance of FireFly-S, we conduct a series of experiments using various SNN models, specifically SCNN5, SCNN7, and SCNN9 models with 5, 7, and 9 convolutional layers. Maintaining consistency with the previous FireFly iteration, we utilized LIF neurons in all experiments, with a time step of 4. Detailed configurations of these models are presented in Table \ref{tab:models}, where 'c3p1' indicates a 3x3 convolution kernel with padding of 1, 'mp2' represents max pooling with a 2x2 filter and stride of 2, and '10fc' represents a fully connected layer with 10 output neurons. These models are trained on the MNIST, DVS-Gesture, and CIFAR-10 datasets, respectively. For testing, all models are deployed on the Xilinx Zynq Ultrascale+ FPGA, utilizing the KV260 Vision AI Starter Kit, which is specifically designed for the deployment of edge-based applications.

On the software side, FireFly-S leverages the Brain-Inspired Cognitive Engine (BrainCog) \cite{zeng2023braincog} and represents a substantial progression in the software-hardware co-design endeavors within the BrainCog project \cite{braincog2024}. The SNN models used in this project are trained using the comprehensive infrastructure provided by BrainCog, ensuring robust and efficient performance metrics that enhance the system's overall functionality.

On the hardware side, FireFly-S is developed using SpinalHDL, and the corresponding Verilog code is produced by the SpinalHDL compiler. These Verilog files are then synthesized and implemented via Xilinx Vivado 2023.2, and the power consumption estimates are derived from the reports generated by the Vivado Design Suite.

\subsection{Parallelism Setup} \label{subsec:parallelism}

\begin{table}[!t]
  \renewcommand\arraystretch{1.25}
  \caption{Parallelism Levels for SCNN5, SCNN7, SCNN9 Models\label{tab:parallelism}}
  \centering
  \begin{tabular}{c|c}
  \hline
  Model & Parallelism $P_{C_o}$\\
  \hline
  \hline
  SCNN5 & (14, 25, 20, 14, 5) \\
  \hline
  SCNN7 & (19, 23, 32, 16, 32, 19, 1) \\
  \hline
  SCNN9 & (32, 16, 21, 13, 26, 13, 20, 10, 1) \\
  \hline
  \end{tabular}
\end{table}

\begin{table}[!t]
  \centering
  \renewcommand\arraystretch{1.25}
  \caption{Resource Usage of FireFly-S on Xilinx Ultrascale KV260}
  \begin{tabular}{c|c|c|c|c}
  \hline
  Model & LUTs & FFs & CARRY8s & BRAMs \\
  \hline
  \hline
  SCNN5 & 30911(26\%) & 55986(24\%) & 268(2\%) & 5(3\%) \\ \hline
  SCNN7 & 55234(47\%) & 98731(42\%) & 473(3\%) & 11(8\%) \\ \hline
  SCNN9 & 84641(72\%) & 139780(60\%) & 1545(11\%) & 76.5(53\%) \\
  \hline
  \end{tabular}
  \label{tab:resources}
\end{table}

\begin{table}[!t]
  \centering
  \renewcommand\arraystretch{1.25}
  \caption{Detailed Resource Usage by Module \protect\\ in the Sixth Convolution Layer of SCNN9 on FireFly-S}
  \begin{tabular}{c|c|c|c|c}
  \hline
  Module & LUTs & FFs & CARRY8s & BRAMs \\
  \hline
  \hline
  Padding Module & 35 & 20 & 0 & 0\\ \hline
  Dataflow Orchestrator & 243 & 258 & 13 & 0.5\\ \hline
  Dual-side Sparsity Detector & 296 & 507 & 9 & 6\\ \hline
  Maxpooling Module & 387 & 612 & 0 & 0\\
  \hline
  \end{tabular}
  \label{tab:modules}
\end{table}

As outlined in Section \ref{sec:architecture}, FireFly-S leverages parallelism to enhance throughput and computational efficiency. Optimizing the parallelism levels for the inter-layer pipelined architecture is crucial to maximizing performance within device constraints, which is achieved through a two-step process.

The first step focuses on balancing the workload across all pipeline stages. This balance is essential to ensure uniform processing times, preventing bottlenecks that could hinder overall performance. We quantify the workload of each stage using the number of multiplex-accumulate (MAC) operations, factoring in the sparsity of computations. The MAC count, representing the total number of algebraic operations required for a convolution layer, is estimated as follows:
\begin{equation}
  \text{MAC} \propto C_o \times C_i \times F_{w_o} \times F_{h_o} \times K_w \times K_h \times T
\end{equation}

Sparsity significantly impacts the actual computational load. We assess sparsity in two stages. First, we calculate the spike sparsity ($s_1$) by inferencing several batches of random images. Second, we determine the weight sparsity ($s_2$). The overall sparsity affecting the workload is the product of these two factors. Given that each stage in our pipeline has $P_{C_o}$ parallel units, the estimated workload for each unit becomes:
\begin{equation}
  \text{Workload per Unit} = \text{MAC} \times \frac{s_1 \times s_2}{P_{C_o}}. 
\end{equation}

By adjusting the parallelism levels, we ensure an even workload distribution across all pipeline stages. In the second step, we gradually increase the parallelism levels until one of two conditions is met: either the available hardware resources, such as Look-Up Tables (LUTs) or RAM, become the limiting factor, or the system achieves the desired performance speed. This iterative adjustment of the parallelism configuration ensures that the pipeline stages are optimized for maximum throughput and efficiency within the given resource constraints. Table \ref{tab:parallelism} presents the setup of parallelism levels for the SCNN5, SCNN7, SCNN9 models in our experiments.

\subsection{Resource and Performance Evaluation}

\begin{table*}
  \centering
  \renewcommand\arraystretch{1.25}
  \begin{threeparttable}[b]
  \caption{Comparison with Related Work for Multiple Image Classification Tasks Using SNNs for Multiple Dataset.}
  \begin{tabular}{c|c|c|c|c|c|c|c|c|c}
  \hline
  Work & Network & Dataset & Acc. & Prec. & FPS & Power & FPS/W & Freq. & Device \\
  \hline
  \hline
  Minitaur \cite{neil2014minitaur} & 784-500-500-10 & MNIST & 92.00 & 16 & 108 & 1.500 & 72.00 & 75 & xc6slx150t \\ \hline

  Han \textit{et al.} \cite{han2020hardware} & 784-1024-1024-10 & MNIST & 97.06 & 16 & 161 & 0.477 & 337.53 & 200 & xc7z045 \\ \hline

  Zhang \textit{et al.} \cite{zhang2019asynchronous} & 784-512-384-10 & MNIST & 98.00 & 8 & 909 & 0.700 & 1298.57 & - & xc7vx690t \\ \hline

  Ju \textit{et al.} \cite{ju2020fpga} & 28x28-64c5-p2-64c5-p2-128-10 & MNIST & 98.94 & 8 & 164 & 4.600 & 35.65 & 150 & xczu9eg \\ \hline

  Fang \textit{et al.} \cite{fang2020encoding} & 28x28-32c3-p2-32c3-p2-256-10 & MNIST & 99.20 & 16 & 133 & 4.500 & 29.55 & 125 & xczu9eg \\ \hline

  \multirow{3}{*}{Ye \textit{et al.} \cite{ye2022implementation}} & \multirow{3}{*}{32x32-32c3-p2-32c3-p2-256-10} & MNIST & 99.10 & \multirow{3}{*}{16} & 820 & \multirow{3}{*}{0.982} & 835.03 & \multirow{3}{*}{200} & \multirow{3}{*}{xc7k325t} \\ \cline{3-4} \cline{6-6} \cline{8-8}
   &  & FMNIST & 90.29 &  & 833 &  & 848.27 &  &  \\ \cline{3-4} \cline{6-6} \cline{8-8}
   &  & SVHN & 82.15 &  & 826 &  & 841.14 &  \\ \hline


  \multirow{3}{*}{E3NE \cite{gerlinghoff2021e3ne}} & LeNet5 & MNIST & 99.30 & 3 & 3400 & 3.400 & 1000.00 & 200 & \multirow{3}{*}{xcvu13p} \\ \cline{2-9}
    & AlexNet & CIFAR10 & 80.60 & \multirow{2}{*}{6} & 14.3 & 4.700 & 3.04 & \multirow{2}{*}{150} &  \\ \cline{2-4} \cline{6-8}
    & VGG11 & CIFAR100 & 65.00 &  & 6.1 & 5.000 & 1.22 &  &  \\ \hline
  
  \multirow{3}{*}{SyncNN \cite{panchapakesan2022syncnn}} & LeNet-L & MNIST & 99.60 & 4 & 1629 & \multirow{3}{*}{0.400\tnote{1}} & 4072.50 & \multirow{3}{*}{200} & \multirow{3}{*}{xczu9eg} \\ \cline{2-6} \cline{8-8}
    & NIN & CIFAR10 & 88.09 & \multirow{2}{*}{8} & 147 &  & 367.5 &  &  \\ \cline{2-4} \cline{6-6} \cline{8-8}
    & VGG13 & SVHN & 95.65 &  & 65 &  & 162.50 &  &  \\ \hline
  
  \multirow{3}{*}{FireFly \cite{li2023firefly}} & 28x28-16c3-64c3-p2-128c3- & \multirow{2}{*}{MNIST} & \multirow{2}{*}{98.12} & \multirow{6}{*}{8} & 2036 & \multirow{3}{*}{3.100} & 656.78 & \multirow{3}{*}{300} & \multirow{3}{*}{xczu3eg} \\ \cline{6-6} \cline{8-8}
    & p2-256c3-256c3-10 &  &  &  & 4975 &  & 1015.31 &  &  \\ \cline{2-4} \cline{6-6} \cline{8-8}
    & 32x32-16c3-64c3-p2-128c3-128c3- & \multirow{2}{*}{CIFAR10} & \multirow{2}{*}{91.36} &  & 966 &  & 311.61 &  &  \\ \cline{1-1} \cline{6-10}
  \multirow{3}{*}{FireFly v2 \cite{li2024firefly}} & p2-256c3-256c3-p2-512c3-10 &  &  &  & 2342 & \multirow{3}{*}{4.900} & 477.95 & \multirow{3}{*}{500\tnote{2}} & \multirow{3}{*}{xczu5ev} \\ \cline{2-4} \cline{6-6} \cline{8-8}
    & 48x48-16c3-64c3-64c3-p2-128c3-128c3- & \multirow{2}{*}{DVS-Gesture} & \multirow{2}{*}{89.29} &  & 282 &  & 90.97 &  & \\ \cline{6-6} \cline{8-8}
    & p2-256c3-256c3-p2-512c3-512c3-10 &  &  &  & 781 &  & 159.39 &  &  \\ \hline

  \multirow{3}{*}{\textbf{Ours}} & \textbf{SCNN5} & \textbf{MNIST} & \textbf{98.03} & \multirow{3}{*}{\textbf{4}} & \textbf{10047} & \textbf{1.301} & \textbf{7722.52} & \multirow{3}{*}{\textbf{333}} & \multirow{3}{*}{\textbf{xczu5ev}} \\ \cline{2-4} \cline{6-8}
    & \textbf{SCNN7} & \textbf{DVS-Gesture} & \textbf{92.05} &  & \textbf{3683} & \textbf{1.821} & \textbf{2022.52} &  &  \\ \cline{2-4} \cline{6-8}
    & \textbf{SCNN9} & \textbf{CIFAR10} & \textbf{87.00} &  & \textbf{2327} & \textbf{3.799} & \textbf{612.53} &  &  \\ \hline

  \end{tabular}
  \label{tab:performance}
  \begin{tablenotes}
  \item[1] SyncNN calculates power usage by measuring the active power consumption of the FPGA board (24.5 W) and subtracting the static power measured when the board is idle (24.1 W). By contrast, it is believed that most existing research acquires power metrics directly from the Vivado report.
  \item[2] The base clock frequency of FireFly v2 is 250 MHz, with a DSP double data rate clock of 500 MHz.
\end{tablenotes}
\end{threeparttable}
\end{table*}

FireFly-S demonstrates its versatility by efficiently training and deploying various SNN models tailored to distinct image classification tasks. This evaluation includes three advanced SNN architectures—SCNN5, SCNN7, and SCNN9—tested across three diverse datasets. These datasets encompass both traditional static images and dynamic, neuromorphic data streams, each presenting unique classification challenges.

Table \ref{tab:resources} details the resource utilization of FireFly-S across various model deployments on the Xilinx Ultrascale KV260, equipped with an xczu5ev chip. As model sizes increase, we observe a corresponding rise in the utilization of the resources. This trend is attributable to an increase in the number of layers and the characteristics of spatial architectures. Notably, the efficient 4-bit quantization and the intrinsic spike-based operation of the SNNs allow inference processes to operate without any DSP Slices, relying solely on LUTs. Additionally, the application of model size compression through 4-bit quantization and high sparsity levels significantly reduces BRAM usage. For example, although the 8-bit quantized weight parameters of SCNN9 still require 9.5Mb—surpassing the chip's BRAM capacity of 5.1Mb and 3.5Mb of Distributed RAM—the reduction techniques lower the BRAM demand to just 53\%. This enhanced efficiency supports the feasible deployment of larger models on the chip.

Table \ref{tab:modules} offers a detailed breakdown of resource usage by each module within a single layer of FireFly-S, indicating that the Dataflow Orchestrator and Dual-side Sparsity Detector are the primary consumers of resources when the Maxpooling module is inactive. Note that the reported resource usage for the Dual-side Sparsity Detector, except for BRAMs, is calculated by dividing by the parallelism level (26 in this experiment). Therefore, while increasing parallelism in FireFly-S can enhance performance, it will also raise the consumption of LUTs and FFs, potentially reaching the resource limits.

In terms of performance, the evaluation uses the Frames Per Second per Watt (FPS/W) metric to assess energy efficiency, recognizing the challenge in comparing studies with varying benchmark complexities. FPS/W comparisons are particularly informative when the models being compared have similar classification accuracies, ensuring a fair assessment of both energy efficiency and algorithmic optimization at a given performance level.

Table \ref{tab:performance} demonstrates that FireFly-S adeptly adjusts to a variety of datasets and models, delivering accuracies on par with those documented in other studies while operating at increased frequencies due to its fully pipelined inter-layer architecture. Additionally, FireFly-S showcases notable energy efficiency. When compared with E3NE, it achieves $\times$7.7 higher FPS/W on MNIST, $\times$202 higher FPS/W on CIFAR10, and a significant 6.4\% improvement in accuracy. Against SyncNN, FireFly-S attains $\times$1.9 higher FPS/W on MNIST and $\times$1.7 higher FPS/W on CIFAR10, despite SyncNN only accounting for dynamic power. These enhancements are supported by two critical advancements: algorithmic improvements that allow for the deployment of large, compressed models on resource-limited FPGAs without sacrificing accuracy, and the incorporation of parallel Dual-side Sparsity Detectors in each layer to exploit network sparsity and boost throughput.

Furthermore, Table \ref{tab:performance} highlights prevalent data precisions in SNN implementations, specifically 16-bit and 8-bit. While the study by Gerlinghoff et al. \cite{gerlinghoff2021e3ne} applies 3-bit quantization to LeNet5, it utilizes 6-bit precision for more complex models like AlexNet and VGG11 to maintain acceptable accuracy levels. In a similar vein, Panchapakesan et al. \cite{panchapakesan2022syncnn} implement percentile-based quantization in SyncNN, achieving 4-bit precision for LeNet-L and 8-bit for NIN and VGG13.However, applying 4-bit quantization to NIN and VGG13 resulted in significant accuracy reductions. As shown in Table \ref{tab:syncnn}, while SyncNN maintains relatively high accuracy with 4-bit quantization for LeNet-L, more complex models like VGG13 and NIN experience substantial accuracy drops, from 95.65\% to 87.87\% and from 88.09\% to 80.33\%, respectively.

In contrast, FireFly-S tends to maintain accuracy after applying 4-bit quantization, especially in more complex datasets like CIFAR10, where the accuracy drop is minimized (from 87.86\% to 87.15\% for SCNN9). These enhanced accuracy levels are primarily due to our quantization-aware and pruning-aware training strategy, which leverages learnable per-channel scales and a hardware-friendly, computation-error-aware approach. In some cases, it may even result in slight accuracy improvements, as seen with SCNN7, where accuracy increases from 87.12\% to 87.87\%. In addition to experiencing less accuracy drop compared to SyncNN with our quantization method, we also combine it with pruning to further compress the network. Specifically, we achieve sparsity levels of 93.82\% for SCNN5, 93.26\% for SCNN7, and 86.12\% for SCNN9, while maintaining competitive accuracy loss. For example, SCNN9 experiences only a slight accuracy drop from 87.86\% to 87.00\% after both quantization and pruning. Notably, in the DVS-Gesture dataset, SCNN7 even shows an accuracy improvement, rising from 87.12\% (without compression) to 92.05\% after applying our compression techniques. These results highlight that our method not only offers a hardware-friendly approach but also achieves highly competitive compression rates while preserving accuracy, demonstrating our joint quantization and pruning strategy is well-suited for low-precision and hardware-optimized inference.

\begin{table}[!t]
  \centering
  \renewcommand\arraystretch{1.25}
  \begin{threeparttable}[b]
  \caption{Compression and Accuracy Results Compared to SyncNN}
  \begin{tabular}{c|c|c|c}
  \hline
  \multirow{2}{*}{SyncNN} & LeNet-L & VGG13 & NIN \\
   & Acc.@MNIST & Acc.@SVHN & Acc.@CIFAR10 \\ \hline \hline
  Base\tnote{1} & 99.60\tnote{2} & 95.65 & 88.09 \\ \hline
  Quant & 99.60 & 87.87\tnote{2} & 80.33 \\ \hline \hline
  \multirow{2}{*}{Ours} & SCNN5 & SCNN7 & SCNN9 \\
    & Acc.@MNIST & Acc.@DVSG & Acc.@CIFAR10 \\ \hline \hline
  Base & 99.57 & 87.12 & 87.86 \\ \hline
  Quant & 99.54 & 87.87 & 87.15 \\ \hline
  Prune & ${\text{98.56}}_{\text{@93.19}}$\tnote{3} & ${\text{92.42}}_{\text{@92.62}}$ & ${\text{86.57}}_{\text{@85.74}}$ \\ \hline
  Quant + Prune & ${\text{98.03}}_{\text{@93.82}}$ & ${\text{92.05}}_{\text{@93.26}}$ & ${\text{87.00}}_{\text{@86.12}}$ \\ \hline
  \end{tabular}
  \label{tab:syncnn}
  \begin{tablenotes}
  \item[1] The base accuracy reported for SyncNN refers to 8-bit quantization, not an uncompressed model.
  \item[2] Values are estimates based on Fig. 9 of SyncNN\cite{panchapakesan2022syncnn}, as no exact values were provided in the original text.
  \item[3] The symbol "$\text{X}_{\text{@Y}}$" indicates that accuracy = X\% and sparsity = Y\%.
  \end{tablenotes}
  \end{threeparttable}
\end{table}



Notably, despite the resource constraints of the edge-tier xczu5ev device, FireFly-S still demonstrates significant improvements across all benchmarks. By utilizing a larger device such as the xczu7ev in the ZCU104, the inference performance can be further enhanced due to the substantially greater number of LUT resources available, which alleviates the bottlenecks experienced with the xczu5ev and supports higher levels of parallelism. Additionally, it is noteworthy that our model is fully mapped on-chip, which means that each layer utilizes fewer resources compared to an overlay architecture. Despite this, our design achieves remarkable performance without any off-chip memory access, thereby eliminating external power consumption typically associated with overlay architectures.

\section{Conclusions} \label{sec:conclusion}
In this work, we introduce a highly energy-efficient and reconfigurable hardware accelerator for SNNs, leveraging a software-hardware co-design approach to optimize energy efficiency during SNN inference. On the software side, our approach includes a quantization-aware and pruning-aware training method, which enables the training of SNN models at 4-bit quantization and 85--95\% sparsity with competitive accuracy loss. This training strategy not only allows the deployment of large SNN models on-chip but also enhances calculation logic utilization by introducing weight sparsity through pruning. On the hardware side, we capitalize on the natural sparsity of spikes inherent to SNNs and the induced sparsity of pruned weights to accelerate processing. This is effectively exploited by our Dual-side Sparsity Detector, which accelerates processing by leveraging both types of sparsity and skipping the unnecessary calculations. Additionally, our architecture is designed to be fully pipelined, enhancing throughput. The Dataflow Orchestrator plays a critical role in maintaining consistent dataflow between layers, minimizing stalls, and simplifying logic to reduce resource consumption. We validate the reconfigurability of our FireFly-S accelerator by evaluating various deep SNN models across multiple datasets. Utilizing commercially available FPGA edge devices offers a more practical and feasible solution compared to other existing hardware platforms. Future work will focus on enhancing the SNN hardware infrastructure further, aiming for ultra-high compression rates and addressing efficiency declines at extreme sparsity levels in Bitmap representation.

\bibliographystyle{IEEEtran}

\vspace{-33pt}
\begin{IEEEbiography}
  [{\includegraphics[width=1in,height=1.25in,clip,keepaspectratio]{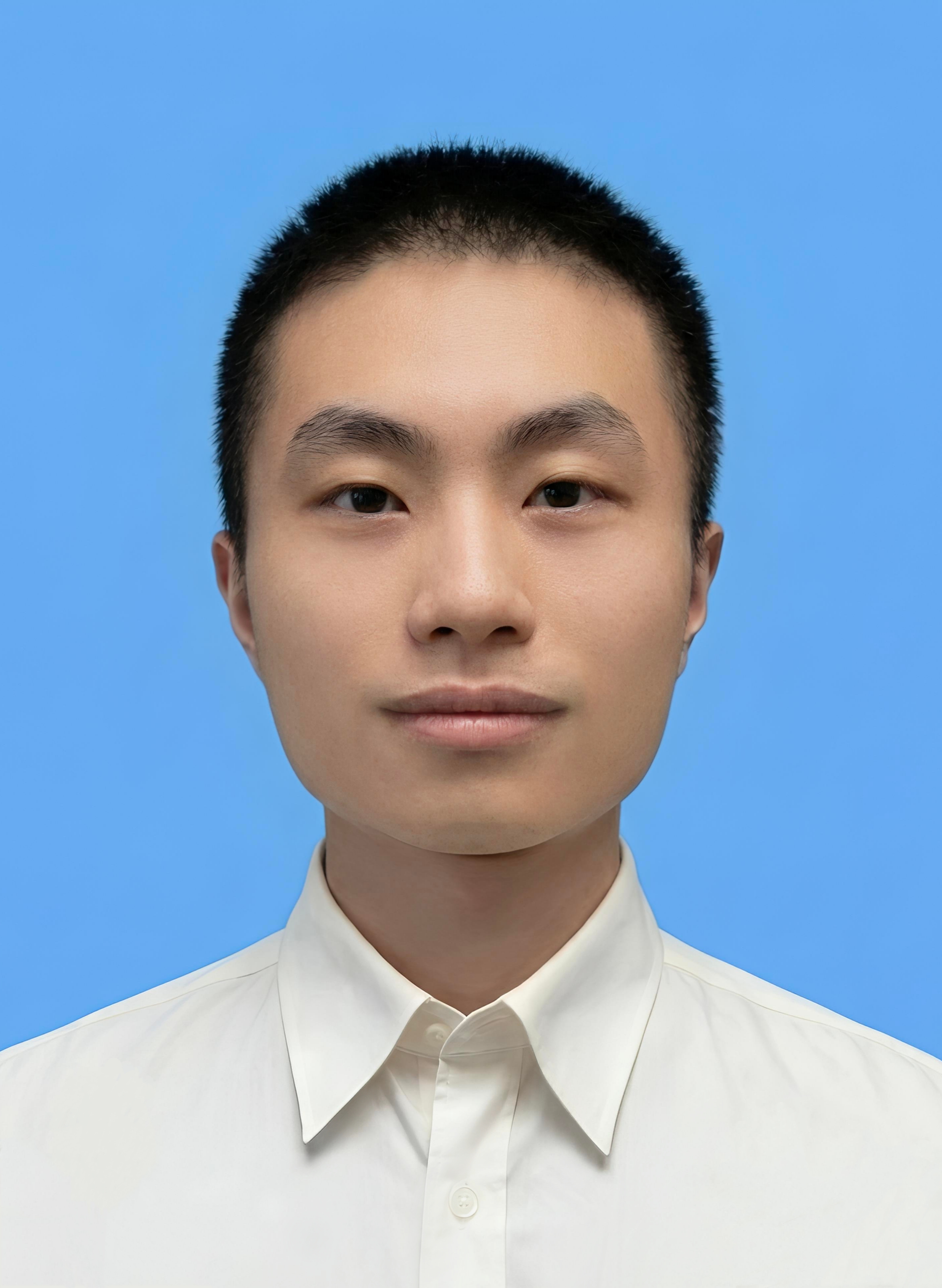}}]
  {Tenglong Li} received his bachelor degree from Sun Yat-sen University in Guangzhou, Guangdong, China in 2023. He is now a master student in the Brain-inspired Cognitive Intelligence Lab, at the Institute of Automation, Chinese Academy of Sciences, under the supervision of Prof. Qian Zhang and Prof. Yi Zeng. His research focuses on hardware acceleration of brain-inspired algorithms, domain-specific architecture and FPGA system design.
\end{IEEEbiography}
\vspace{-33pt}
\begin{IEEEbiography}
  [{\includegraphics[width=1in,height=1.25in,clip,keepaspectratio]{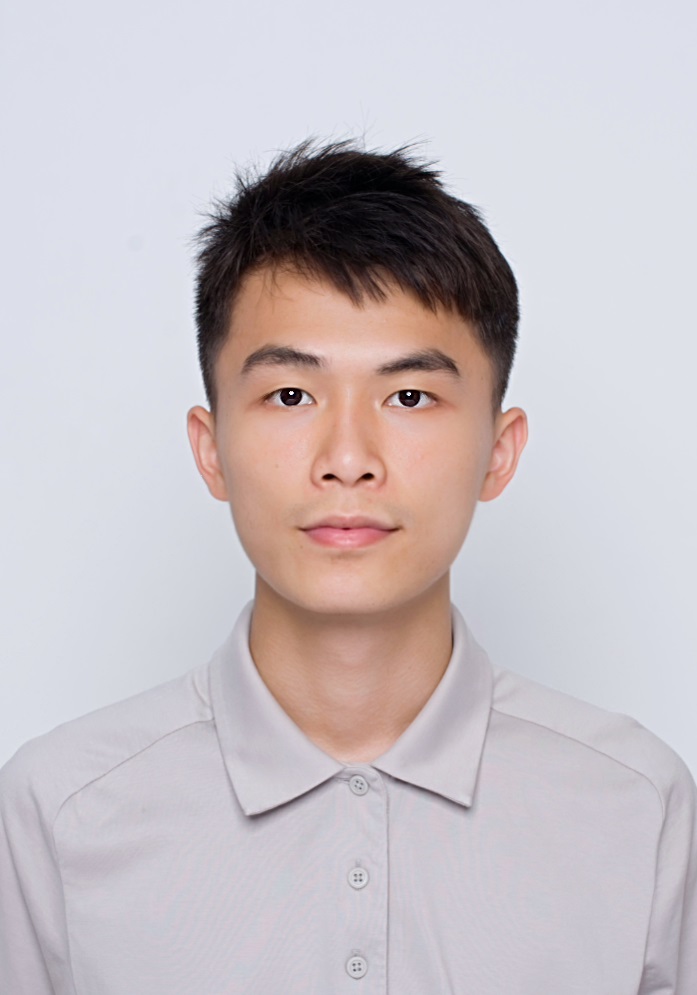}}]
  {Jindong Li} received his bachelor degree from Sun Yat-sen University in Guangzhou, Guangdong, China in 2022. He is now a master student in the Brain-inspired Cognitive Intelligence Lab, at the Institute of Automation, Chinese Academy of Sciences, under the supervision of Prof. Qian Zhang and Prof. Yi Zeng. His research focuses on hardware acceleration of brain-inspired algorithms, domain-specific architecture and FPGA system design.
\end{IEEEbiography}
\vspace{-33pt}
\begin{IEEEbiography}
  [{\includegraphics[width=1in,height=1.25in,clip,keepaspectratio]{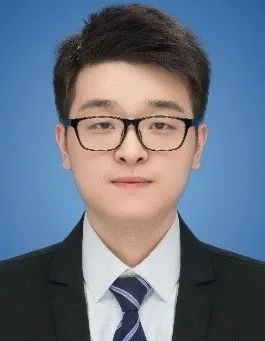}}]
  {Guobin Shen} received his bachelor degree from Sun Yat-sen University in Guangzhou, Guangdong, China in 2021. He is now a PhD candidate in the Brain-inspired Cognitive Intelligence Lab, at the Institute of Automation, Chinese Academy of Sciences, under the supervision of Prof. Yi Zeng. His research focuses on biologically-inspired learning algorithms and spiking neural network architecture design and training strategies.
\end{IEEEbiography}
\vspace{-33pt}
\begin{IEEEbiography}
  [{\includegraphics[width=1in,height=1.25in,clip,keepaspectratio]{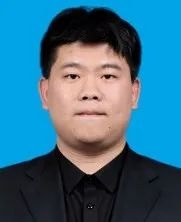}}]
  {Dongcheng Zhao} received the bachelor degree from XiDian University, Xi'an, Shaanxi, China, in 2016 and Ph.D degree from University of Chinese Academy of Sciences, Beijing, China, in 2021. He is currently an assistant professor in the Brain-inspired Cognitive Intelligence Lab, Institute of Automation, Chinese Academy of Sciences, China. His current research interests include learning algorithms in spiking neural networks, thalamus-cortex interaction, visual object tracking, etc.
\end{IEEEbiography}
\vspace{-33pt}
\begin{IEEEbiography}
  [{\includegraphics[width=1in,height=1.25in,clip,keepaspectratio]{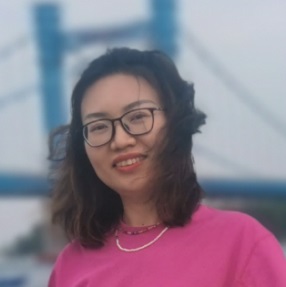}}]
  {Qian Zhang} obtained her bachelor degree in 2009 and Ph.D degree in 2014 from Xidian University, China. She is currently an associate professor and director in the Brain-inspired Cognitive Intelligence Lab, Institute of Automation, Chinese Academy of Sciences, China. Her research interests include brain simulation and brain-inspired cognitive computing modeling, especially working memory modeling and simulation of brain rhythms at different levels of consciousness.
\end{IEEEbiography}
\vspace{-33pt}
\begin{IEEEbiography}
  [{\includegraphics[width=1in,height=1.25in,clip,keepaspectratio]{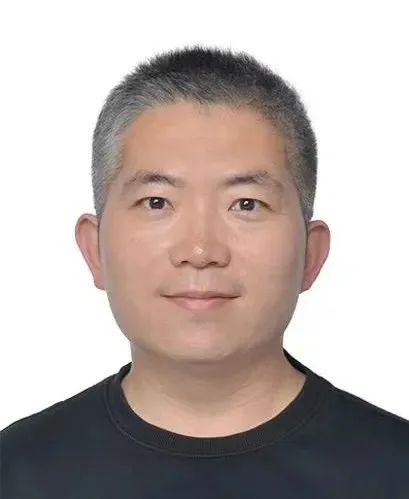}}]
  {Yi Zeng} obtained his bachelor degree in 2004 and Ph.D degree in 2010 from Beijing University of Technology, China. He is currently a professor and director in the Brain-inspired Cognitive Intelligence Lab, Institute of Automation, Chinese Academy of Sciences, China. He is a principal investigator in the Center for Excellence in Brain Science and Intelligence Technology, Chinese Academy of Sciences, China, and a Professor in the School of Future Technology, and School of Humanities, University of Chinese Academy of Sciences, China. His research interests include brain-inspired artificial intelligence, brain-inspired cognitive robotics, ethics and governance of Artificial Intelligence, etc.
\end{IEEEbiography}


 





\end{document}